\documentclass[prd,aps,reprint,groupedaddress,superscriptaddress,nofootinbib,twocolumn]{revtex4}
\usepackage{graphicx}
\usepackage{amsmath,amssymb,bm}
\usepackage{hyperref}
\usepackage{autobreak,mathtools}
\usepackage{comment}
\usepackage[dvipsnames, usenames]{xcolor}
\usepackage[scr=boondoxo,scrscaled=1.05]{mathalfa}

\definecolor{p-r}{RGB}{171, 40, 52}
\hypersetup{colorlinks=true, citecolor=p-r, linkcolor=p-r,urlcolor=p-r}

\begin{document}

\title{Charging up regular black holes}
\author{Ra\'ul Carballo-Rubio}
\affiliation{Instituto de Astrof\'isica de Andaluc\'ia (IAA-CSIC),
Glorieta de la Astronom\'ia, 18008 Granada, Spain}
\author{Chiara Coviello}
\affiliation{Department of Physics, King’s College London, The Strand, London, WC2R 2LS, UK}
\author{Vania Vellucci}
\affiliation{Quantum Theory Center (${\hbar}$QTC) \& D-IAS, IMADA at Southern Denmark Univ., Campusvej 55, 5230 Odense M, Denmark}

\begin{abstract}
We present a general construction of charged regular black holes as solutions of a generalization of the Einstein--Maxwell field equations in spherical symmetry in which the Einstein tensor is deformed into an identically conserved tensor containing up to second derivatives of the gravitational field. The generality of the construction allows us to define the field equations satisfied by generic regular black holes when becoming charged. The conditions that guarantee regularity of charged solutions are evaluated and shown to be more stringent than the regularity conditions for uncharged solutions. This implies, in particular, that the charged versions of the Bardeen and Hayward black holes become singular. Improved versions of the Bardeen and Hayward metrics that remain regular when charged are proposed. Our results indicate that regularizing the vacuum solutions of general relativity is, in general, not enough to yield regular solutions in other situations of physical interest. The implications that follow for the construction of realistic regular black holes, in which aspects such as rotation and the presence of matter fields are taken into account, are discussed.
\end{abstract}

\maketitle

\section{Introduction}

Regular black holes are theoretical alternatives to the Kerr family of vacuum solutions of the Einstein field equations~\cite{Kerr:1963ud,Visser:2007fj}, defined with the goal of alleviating their singular nature~\cite{Carballo-Rubio:2019fnb}. The study of regular black holes is an active area of research~\cite{Carballo-Rubio:2025fnc}. Regular black holes were first described as geometric deformations of the Schwarzschild black hole by Bardeen~\cite{Bardeen1968}, and further explored as possible outcomes of physics beyond general relativity by Frolov, Vilkovisky, Dymnikova and Hayward~\cite{Frolov:1981mz,Dymnikova:1992ux,Hayward:2005gi,Frolov:2014jva,Dymnikova:2015yma}.

The physical viability of regular black holes as consistent and predictive models has remained unclear for decades~\cite{Carballo-Rubio:2018pmi}, in particular because of a lack of understanding of their dynamical properties. Key open issues were describing the formation of regular black holes from the collapse of standard matter and the backreaction of matter perturbations on regular black hole interiors~\cite{Brown:2011tv,Frolov:2017rjz,Carballo-Rubio:2021bpr,Carballo-Rubio:2024dca}. 

Describing gravitational collapse to regular black holes is possible in a phenomenological approach first explored by Ziprick and Kunstatter and valid for spherically symmetric spacetimes~\cite{Ziprick:2010vb,Kunstatter:2015vxa,Boyanov:2025pes}. Theories in this approach are characterized by the deformation of the spherically symmetric Einstein tensor into an identically conserved tensor~\cite{Carballo-Rubio:2025ntd}. Deforming the Einstein tensor provides a mechanism to regularize the entire manifold of vacuum solutions of black holes characterized by a single parameter, their mass $M$ (thus solving some issues that arise when regularization is achieved by deforming the action of matter fields, in particular the coexistence of regular and singular solutions in non-linear electrodynamics~\cite{Huang:2025uhv}).

An independent research line exploring higher-derivative gravitational theories known as quasitopological gravities~\cite{Myers:2010ru,Oliva:2010eb,Bueno:2016xff,Hennigar:2017ego,Ahmed:2017jod,Bueno:2019ycr}, defined for $D\geq5$ and displaying second-order field equations in spherical symmetry, resulted in the description of gravitational collapse to regular black holes~\cite{Bueno:2024eig,Bueno:2024zsx,Bueno:2025gjg}. These solutions are naturally included in the phenomenological approach described in the previous paragraph, which is a remarkable convergence of research programs with different scope and goals. In $D=4$ dimensions, gravitational theories with second-order field equations in spherical symmetry and regular black hole solutions still exist, but must be non-polynomial~\cite{Bueno:2025zaj,Borissova:2026wmn,Borissova:2026krh}.

In general relativity, vacuum black holes are characterized by their mass $M$, charge $Q$ and angular momentum $J$. Regularizing black holes with non-zero charge and angular momentum is a key open issue in the literature. While situations including rotation are the most interesting ones from the perspective of realistic modeling, the study of charged situations is an important intermediate step that, while simpler mathematically, can shed light into the features and challenges that arise when regularizing rotating black holes.

\section{Field equations}

A spherically symmetric gravitational metric $g_{\mu\nu}$ defined on a manifold $\mathcal{M}=\mathcal{N}\times S^2$ can be parametrized in the ``warped product'' form~\cite{ONeillbook}:
\begin{equation}
g_{\mu\nu}(y)\text{d}y^\mu\text{d}y^\nu=q_{ab}(x)\text{d}x^a\text{d}x^b+r^2(x)\gamma_{ij}\text{d}\theta^i\text{d}\theta^j,    
\end{equation}
where $q_{ab}(x)$ and $r(x)$ are, respectively, a 2-dimensional metric and a scalar field on $\mathcal{N}$, and $\gamma_{ij}$ is the standard metric on $S^2$.

The most general field equations for $g_{\mu\nu}$ above, in which the spherically symmetric Einstein tensor is deformed into an identically conserved tensor constructed from up second derivatives of $q_{ab}(x)$ and $r(x)$, take the form~\cite{Carballo-Rubio:2025ntd}:
\begin{equation}\label{eq:mfe}
\mathscr{G}_{\mu\nu}(q,r)=8\pi T_{\mu\nu},\qquad 
\end{equation}
with the left-hand side written as
\begin{equation}\label{eq:calgdef}
\mathscr{G}_{\mu\nu}(q,r)=\frac{1}{r^2}\mathscr{E}_{ab}\delta^a_\mu\delta^b_\nu-\frac{1}{4}r\mathscr{F}\gamma_{ij}\delta^i_\mu\delta^j_\nu,
\end{equation}
where
\begin{align}\label{eq:geneqs1}
\mathscr{E}_{ab}(q,r)&=\bm{\beta}\nabla_a\nabla_b r-q_{ab}\left(\frac{1}{2}\bm{\alpha}+\bm{\beta}\square r \right)+\bm{\gamma}\nabla_ar\nabla_br
,\\
\mathscr{F}(q,r)&=-\bm{\beta}\mathcal{R}+2\partial_r\bm{\beta}\square r+\partial_r\bm{\alpha}\nonumber\\
&+2\partial_\chi\bm{\beta}\left[\left(\square r\right)^2-\nabla_a\nabla_b r\nabla^a\nabla^br\right]\nonumber\\
&-2\partial_r\bm{\gamma}\chi-2\bm{\gamma}\square r-2\partial_\chi\bm{\gamma}\nabla_a r\nabla^a\chi,\label{eq:geneqs2}
\end{align}
where we have defined $\bm{\gamma}=\partial_\chi\bm{\alpha}-\partial_r\bm{\beta}$, $\chi=\nabla_a r\nabla^a r$ and $\mathcal{R}$ as the Ricci scalar of $q_{ab}(x)$. The functions $\bm \alpha(r,\chi)$ and $\bm \beta(r,\chi)$ define the specific theory under consideration, and we require them to recover the general relativity limit in the asymptotic region $r \rightarrow \infty$.

The field equations~\eqref{eq:mfe} describe a wide range of modifications of general relativity which include, in particular, modifications in which black hole solutions have bounded curvature both in vacuum and in the presence of matter~\cite{Carballo-Rubio:2025ntd,Boyanov:2025pes}.

The main novelty in this paper is the study of electrovacuum solutions with electromagnetic potential
\begin{equation}\label{eq:adef}
A_\mu(y)\text{d}y^\mu=A_a(x)\text{d}x^a.    
\end{equation}
This potential satisfies the Maxwell equations
\begin{equation}\label{eq:maxeqs}
\nabla_\nu F^{\nu\mu}=\frac{1}{\sqrt{-g}}\partial_\nu\left(\sqrt{-g}F^{\nu\mu}\right)=4\pi J^\mu,    
\end{equation}
with source 
\begin{equation}\label{eq:maxeqssource}
J^t=Q\delta(r),    
\end{equation}
as well as the master field equations~\eqref{eq:mfe} with the usual stress-energy tensor
\begin{equation}\label{eq:maxset}
T_{\mu\nu}=\frac{1}{4\pi}\left(F_{\mu\alpha}F_\nu^{\ \alpha}-\frac{1}{4}g_{\mu\nu}F_{\alpha\beta}F^{\alpha\beta}\right).    
\end{equation}

Note that these equations contain up to second derivatives of the gravitational field but this does not guarantee well-posedness~\cite{Papallo:2017qvl,Papallo:2017ddx}. The appearance of terms like $(\square r)^2$ and $\nabla_a\nabla_b r\,\nabla^a\nabla^b r$ in Eq.~\eqref{eq:geneqs2} might actually seem problematic for well-posedness. However, taking the trace of Eq.~\eqref{eq:geneqs1} provides an expression for $\square r$ and thus also for $\nabla_a\nabla_b r$ in terms of first-order derivatives of $r$. Consequently, only ${\cal R}$ introduces second derivatives of $q_{ab}$ on the left-hand side of Eq.~\eqref{eq:geneqs2}. This indicates that, with suitable coordinate conditions, the system can be cast as a well-posed problem.

\section{Vacuum solutions}

We start with a brief review of vacuum solutions, that is, with vanishing electromagnetic fields. These were studied in~\cite{Carballo-Rubio:2025ntd} in the notation used here, and in~\cite{Kunstatter:2015vxa} for an equivalent vacuum theory (a subspace of vacuum solutions can describe renormalization-group improved black holes, as discussed in~\cite{Borissova:2026dlz}).

The deformation of the Einstein tensor in terms of $\bm{\alpha}$ and $\bm{\beta}$ in Eq.~\eqref{eq:mfe} induces a deformation of vacuum solutions. Considering
\begin{equation}
    ds^2= -h(r)dt^2+ \frac{dr^2}{f(r)}+ r^2 d\Omega^2,
    \label{eq:metric_back}
\end{equation}
the field equations~\eqref{eq:mfe} are solved by
\begin{equation}\label{eq:vaceqs}
\frac{\text{d}f(r)}{\text{d}r}=-\frac{\bm{\alpha}}{\bm{\beta}},\quad h(r)=f(r)e^{2\int\text{d}r\left.\bm{\gamma}/\bm{\beta}\right|_{\chi=f(r)}}.
\end{equation}
These equations can be used to find the solutions of a specific theory (i.e., fixing $\bm{\alpha}$ and $\bm{\beta}$), or to find theories for which specific metrics arise as vacuum solutions (i.e., fixing $f$ and $h$).

From Eq.~\eqref{eq:vaceqs}, it follows that the subfamily of theories satisfying the integrability condition
\begin{equation}\label{eq:intcond}
\bm{\gamma}=\partial_\chi\bm{\alpha}-\partial_r\bm{\beta}=0    
\end{equation}
is characterized by a single metric function $f(r)$ which, integrating Eq.~\eqref{eq:vaceqs}, is determined by an algebraic relation~\cite{Carballo-Rubio:2025ntd} (see also~\cite{Kunstatter:2015vxa}) in terms of the potential function $\bm \Omega$:
\begin{equation}\label{eq:omegasol}
\left.\bm{\Omega}(r,\chi)\right|_{\chi=f}=4M,    
\end{equation}
where we have normalized this function so that $M$ is the ADM mass, and
\begin{equation}\label{eq:alphabetaomegadef}
\frac{\partial\bm{\Omega}}{\partial r}=\bm{\alpha},\quad \frac{\partial\bm{\Omega}}{\partial \chi}=\bm{\beta}.
\end{equation}

We will focus specifically on two particular deformations of general relativity, though our results are applicable to wider families of theories, for which these two cases serve as specific representatives. 
The first one is the Bardeen metric~\cite{Bardeen:1968xqk},
\begin{equation}\label{eq:bardeen1}
f_{\rm B}(r)=1-\frac{2r^2 M}{\left(r^2+\ell^2\right)^{3/2}},    
\end{equation}
for which
\begin{align}\label{eq:bardeen2}
\bm{\Omega}_{\rm B}(r,\chi)&=2\left(1-\chi\right)\frac{\left(r^2+\ell^2\right)^{3/2}}{r^2},\nonumber\\
\bm{\alpha}_{\rm B}(r,\chi)&=2\left(1-\chi\right)\frac{\sqrt{r^2+\ell^2}\left(r^2-2\ell^2\right)}{r^3},\nonumber\\ \bm{\beta}_{\rm B}(r,\chi)&=-\frac{2\left(r^2+\ell^2\right)^{3/2}}{r^2}.   
\end{align}
This theory is a representative of the Ziprick--Kunstatter class~\cite{Ziprick:2010vb,Boyanov:2025pes}.

The second one is the Hayward metric~\cite{Hayward:2005gi},
\begin{equation}\label{eq:hayward1}
f_{\rm H}(r)=1-\frac{2r^2M}{r^3+2\ell^2 M},   
\end{equation}
for which
\begin{align}\label{eq:hayward2}
\bm{\Omega}_{\rm H}(r,\chi)&=2\left(1-\chi\right)\frac{r^3}{r^2-\ell^2(1-\chi)},\nonumber\\
\bm{\alpha}_{\rm H}(r,\chi)&=\frac{2\left[r^4-3\ell^2r^2\left(1-\chi\right)\right]\left(1-\chi\right)}{\left[r^2-\ell^2\left(1-\chi\right)\right]^2},\nonumber\\
\bm{\beta}_{\rm H}(r,\chi)&=-\frac{2r^5}{\left[r^2-\ell^2\left(1-\chi\right)\right]^2}.   
\end{align}
This theory is a representative of the Kunstatter--Maeda--Taves class~\cite{Kunstatter:2015vxa,Boyanov:2025pes}.

The potential function of general relativity $\bm{\Omega}_{\rm GR}=2(1-\chi)r$ is recovered in both cases in the limit $\ell\rightarrow0$.

For theories not satisfying the integrability condition, the equations above can be solved by introducing an integrating factor~\cite{Kunstatter:2015vxa} (see also~\cite{Borissova:2026dlz}). However, the two examples we will be working with satisfy this condition.

\section{Electrovacuum solutions}

A noteworthy aspect of Eqs.~\eqref{eq:mfe} is that the interaction with matter of spherical black holes beyond general relativity can be studied following the same methodology as in general relativity~\cite{Carballo-Rubio:2025ntd}. We will illustrate this here for electrovacuum solutions.

We now show that the generalization of the Einstein--Maxwell equations given by Eqs.~\eqref{eq:mfe},~\eqref{eq:maxeqs},~\eqref{eq:maxeqssource} and ~\eqref{eq:maxset} can be explicitly integrated without specifying the functions $\bm{\alpha}$ and $\bm{\beta}$ as long as these satisfy the integrability condition~\eqref{eq:intcond} (which has also been noted for vacuum solutions~\cite{Carballo-Rubio:2025ntd} and Vaidya solutions~\cite{Boyanov:2025pes}). We will first write the field equations in complete generality to show that, if the integrability condition is not satisfied, the field equations to be solved become coupled and finding explicit solutions becomes more challenging.

We use the same ansatz for the metric as in Eq.~\eqref{eq:metric_back}, where the only non-zero component of the electromagnetic field~\eqref{eq:adef} is
\begin{equation}
A_t(r)=-\phi(r),\quad F_{tr}(r)=-\partial_r A_t(r)=\phi'(r)=-E_r(r),    
\end{equation}
while the non-zero components of the stress-energy tensor are:
\begin{align}
T_{tt}(r)&=\frac{f(r)}{8\pi}  \left[\phi'(r)\right]^2,\nonumber\\
T_{rr}(r)&=-\frac{1}{8\pi h(r)} \left[\phi'(r)\right]^2,\nonumber\\ 
T_{\theta\theta}(r)&=\frac{f(r)\,r^2 }{8\pi h(r)}\left[\phi'(r)\right]^2,\nonumber\\ 
T_{\varphi\varphi}(r)&=\frac{f(r)\,r^2\sin^2\theta }{8\pi h(r)}\left[\phi'(r)\right]^2.
\end{align}
We have the following three independent field equations:
\begin{align}
\frac{h(r)\left[\bm{\alpha}+\bm{\beta}f'(r)\right]}{2r^2}&=f(r)\left[\phi'(r)\right]^2,\nonumber\\
-\frac{\bm{\alpha}/f(r)+\bm{\beta}h'(r)/h(r)-2\bm{\gamma}}{2r^2}&=-\frac{\left[\phi'(r)\right]^2}{h(r)},\nonumber\\
\frac{\sqrt{f(r) h^{-1}(r)}}{r^2}\partial_r\left[\sqrt{f(r) h^{-1}(r)}r^2 \phi'(r)\right]&=-4\pi Q\delta(r). 
\end{align}
Combining the first two equations results in the following relation, which is the same as in the vacuum case:

\begin{equation}
\partial_r\left[-\log f(r)+\log h(r)\right]=2\frac{\bm{\gamma}}{\bm{\beta}},    
\end{equation}
so that, for theories satisfying $\bm{\gamma}=\partial_\chi\bm{\alpha}-\partial_r\bm{\beta}=0$, we have (up to an irrelevant integration constant)
\begin{equation}\label{eq:int_sol}
    h(r)=f(r).
\end{equation}

In the following, we will restrict our arguments to this subfamily of theories. 

In this case, the field equation for the electromagnetic field reads
\begin{equation}
\frac{1}{r^2}\partial_r\left[ r^2 \phi'(r)\right]=-4\pi Q\delta(r),   
\end{equation}
and is solved by
\begin{equation}
\phi(r)=\frac{Q}{r}.    
\end{equation}
We are thus left with a single equation to solve:
\begin{equation}
\bm{\alpha}+\bm{\beta}f'(r)=\frac{2 Q^2}{r^2}.    
\end{equation}
with $\chi=f(r)$.
As in the uncharged case, this equation is solved by a function $\bm{\Omega}(r,\chi)$, satisfying in this case
\begin{equation}
\left.\bm{\Omega}\left(r,\chi\right)\right|_{\chi=f}=4M-\frac{2Q^2}{r}.    
\end{equation}
This expression generalizes Eq.~\eqref{eq:omegasol} to the charged case.

\begin{figure}[htbp]
\includegraphics[width=\linewidth]{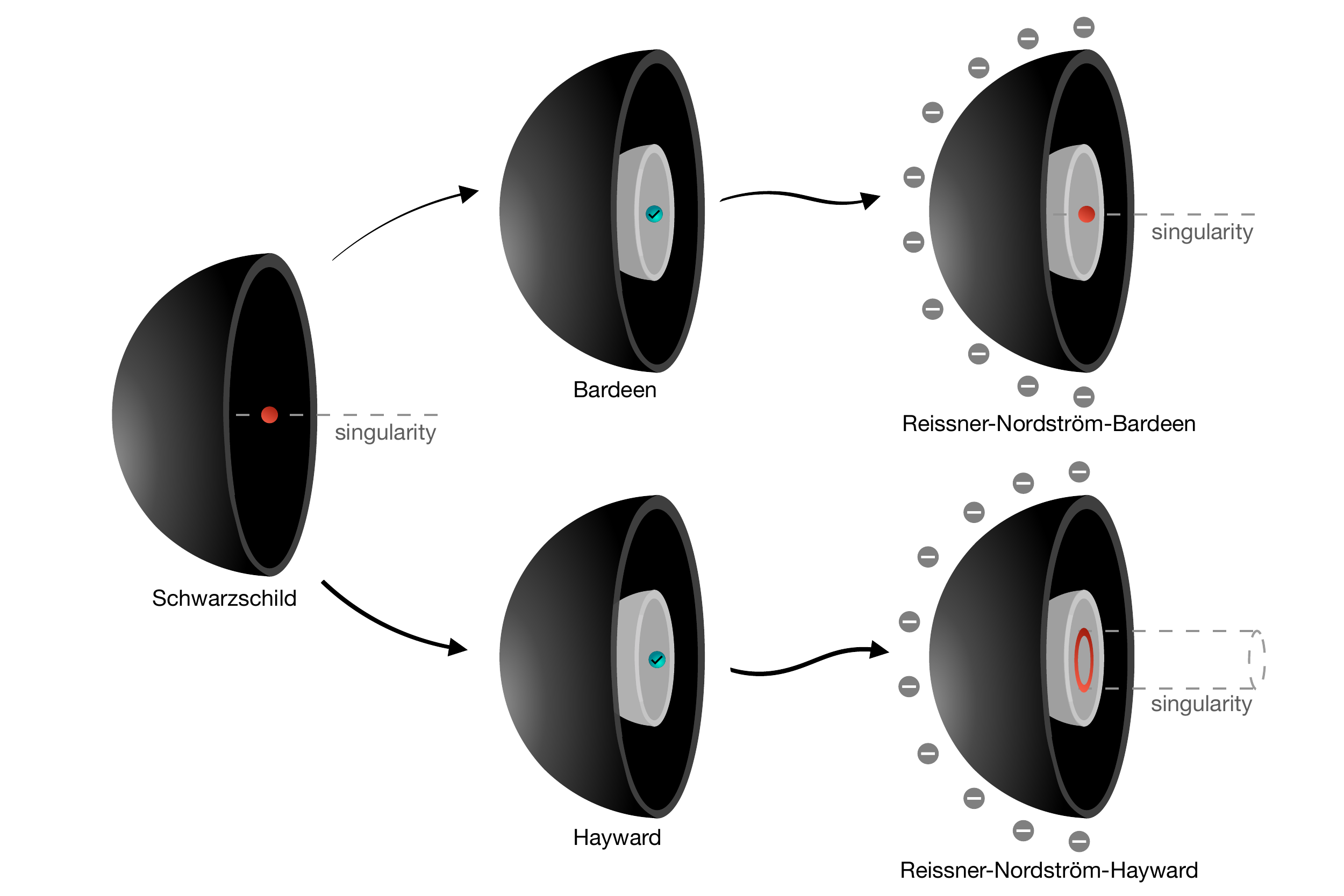}

\caption{The Schwarzschild black hole can be deformed into two separate families of regular black holes, with the Bardeen and Hayward black holes being representatives of each family. These deformations satisfy Eq.~\eqref{eq:mfe} in vacuum. Once charged, the resulting Reissner--Nordstr\"om--Bardeen and Reissner--Nordstr\"om--Hayward black holes become singular at $r=0$ and $r=r_+>0$, respectively. This invites the consideration of two possibilities, depending on whether the regularity of electrovacuum solutions is considered necessary: either choosing different representatives of these families of vacuum solutions that remain regular when charged, or supplementing these solutions with non-vacuum regions with matter contents beyond electrovacuum.\label{fig:diagram}}
\end{figure}

Aside from reproducing the Reissner--Nordstr\"om metric for the general relativity case, this relation allows us to derive charged versions of any regular black hole. In particular, we obtain the generalizations of e.g. the Bardeen and Hayward metrics by recalling Eqs.~(\ref{eq:bardeen1}-\ref{eq:bardeen2}) and~(\ref{eq:hayward1}-\ref{eq:hayward2}), respectively, and solving for the metric function:
\begin{align}
f_{{\rm B},Q}(r)&=1-\frac{r\left(2Mr-Q^2\right)}{\left(r^2+\ell^2\right)^{3/2}} ,\\ f_{{\rm H},Q}(r)&=1-\frac{r^2\left(2Mr-Q^2\right)}{r^4+\ell^2\left(2M r-Q^2\right)}.   
\end{align}
These two metrics are not regular, although in different ways (Fig.~\ref{fig:diagram}).

The charged Bardeen (or Reissner--Nordstr\"om--Bardeen) metric is singular at $r=0$ due to $\left.\partial_r f_{\rm B,Q}\right|_{r=0}=Q^2/\ell^3\neq0$, as the latter vanishing is required for regularity~\cite{Carballo-Rubio:2019fnb}. 

On the other hand, the charged Hayward (or Reissner--Nordstr\"om--Hayward) metric is regular at $r=0$, but it is singular for a positive value of the radius determined by the equation
\begin{equation}
r^4+2 M\ell^2 r - Q^2\ell^2=0,    
\end{equation}
which is guaranteed to have a positive real root $r_+$ due to the intermediate value theorem together with its  behavior in the limits $r\rightarrow\infty$ and $r\rightarrow 0$. This leads to a divergent Kretschmann scalar, as $\lim_{r\rightarrow r_+}\mathcal{K}\propto (r-r_+)^{-4}$.
For a more comprehensive analysis of the conditions required for an electrovacuum regular metric, we refer the reader to~\cite{Carballo-Rubio:2026trz}.

\section{Implications}

We have discussed above how to construct metrics that result from charging up known regular black holes, and shown that this leads to the occurrence of singularities unless some additional conditions are imposed. When faced with similar difficulties, some authors have considered deformations of electrodynamics~\cite{Hennigar:2025yqm,PinedoSoto:2026hfm,2026arXiv260310110B}. The emergence of these additional conditions is actually a general aspect shared by other matter contents such a scalar field~\cite{Carballo-Rubio:2026trz} (see also~\cite{2026arXiv260310110B}), and thus considering deformations of electrodynamics is generally not enough to guarantee regular solutions. A possible solution may be to add an infinite tower of terms involving non-minimal couplings with matter fields, in a similar way as it can be done for electrodynamics ~\cite{2026arXiv260310110B,Cano:2020ezi,Cano:2020qhy}. Here, we focus on solutions in which electrodynamics retains its standard form, and thus rely on either finding other deformations of the gravitational sector that remain regular in the presence of matter, or going beyond purely electrovacuum solutions to construct models of regular black holes.

Our first observation is that, while electrovacuum solutions (in the Einstein--Maxwell sense defined above) are not regular for theories leading to Bardeen and Hayward regular black holes in the uncharged case, other regularizations exist. We discuss below that it is possible to construct generalizations of the Bardeen and Hayward metrics that remain regular when charged (see also the related discussion in~\cite{PinedoSoto:2026hfm}).

For the Bardeen case, a generalization that still belongs to the Ziprick--Kunstatter class and remains regular when charged is given by
\begin{equation}
\bm{\Omega}_{\rm B2.0}(r)=2\left(1-\chi\right)\frac{\left(r^2+\ell^2\right)^{5/2}}{r^4}.
\end{equation}
For the Hayward case, one simple option is replacing
\begin{equation}
r^4+2 M\ell^2 r\rightarrow \sqrt{r^8+(2M)^2\ell^4},    
\end{equation}
which guarantees that the polynomial obtained when replacing $2M\rightarrow 4M-Q^2/r$ has no positive roots. Such a metric results from the following choice in the Kunstatter--Maeda--Taves class:
\begin{equation}
\bm{\Omega}_{\rm H2.0}(r,\chi)=2(1-\chi)\frac{r^4}{\sqrt{r^6-(1-\chi)^2\ell^4}}.    
\end{equation}
This case is similar to the one studied in~\cite{Hao:2025utc}.

We have thus identified specific theories in the Ziprick--Kunstatter and Kunstatter--Maeda--Taves classes that remain regular when charged. In four dimensions, the gravitational actions for these theories can be constructed by following the lifting procedures discussed in~\cite{Borissova:2026wmn,Borissova:2026krh}, in both cases requiring non-polynomial invariants constructed from curvature tensors and, in the former case, their derivatives. 

The requirement of regularity of electrovacuum solutions might be too stringent. Introducing other (static or time-dependent) matter fields aside from electromagnetic fields may still result in regular solutions. This would require a shift in perspective, as regular black holes must include matter in their description for regularity to be attained. We discuss below that it is possible to find regular solutions with non-electrovacuum cores.

Previous results show that the introduction of repulsive forces, either in the form of a non-zero charge~\cite{PoncedeLeon:2017usu} or of a modification of gravity~\cite{Bueno:2025tli,Arrechea:2026ngi}, allows placing a spherically symmetric distribution of a perfect fluid in the interior of regular black holes without adding further (distributional) sources. In such solutions, the perfect fluid acts as the extended source for the gravitational field described by the regular black hole metric. As both repulsive effects act in a synergetic way when combined, the existence of such solutions is guaranteed in the situations analyzed in this paper, as it will be discussed in detail elsewhere. 

\section{Conclusions}

Recent advances have provided a general formalism in which to show that deformations of the gravitational interaction such that the latter becomes weaker at the core of black holes can result in regular vacuum solutions. In this work, we have used this formalism to understand how these vacuum solutions react to the introduction of charge.

We have formulated and solved the most general extension of the Einstein--Maxwell field equations in spherical symmetry in which the Einstein tensor is deformed into an identically conserved tensor containing up to second derivatives of the gravitational field.
Our main technical result is that known families of black holes, such as the Bardeen and Hayward black holes, become singular when charged. We have obtained explicitly the form of the Reissner--Nordstr\"om--Bardeen and Reissner--Nordstr\"om--Hayward black holes and discussed their singularity structure.

The main physical implication of our work is the identification of two avenues along which the construction of regular charged solutions can proceed. The first one is finding other deformations of general relativity which remain regular in the presence of charge, and result in different families of uncharged regular black holes. We have shown that these families exist by providing specific examples. The second one is abandoning the idea of regularizing vacuum or electrovacuum solutions and thinking about regular black holes as describing the exterior to suitable matter distributions. We have discussed that the repulsive forces associated both with deformations of general relativity and the presence of charge allow the construction of such non-electrovacuum metrics by using charged perfect fluids.

We can anticipate that similar issues may arise when considering rotating regular black holes. It is reasonable to expect that deformations of the Bardeen and Hayward metrics incorporating angular momentum, and reducing to these static geometries for $J=0$, may not be regular. It would then be necessary to consider different families of static regular black holes as the starting point, or to introduce rotating sources of matter at the core of these deformations with $J\neq0$. In any case, our results and arguments here are a clear proof that regularizing charged and rotating regular black holes is subtler than regularizing simpler situations and requires new ingredients.

\acknowledgments

We thank Luis Lehner for valuable discussions. R.C-R. acknowledges financial support provided by the Spanish Government through the Ram\'on y Cajal program (contract RYC2023-045894-I), the Grant No.~PID2023-149018NB-C43 funded~by MCIN/AEI/10.13039/501100011033, and the Severo Ochoa grant CEX2021-001131-S funded by MCIN/AEI/ 10.13039/501100011033. CC is supported by King’s College London through an NMES funded studentship.
VV is grateful for the hospitality of Perimeter Institute where part of this work was carried out. Research at Perimeter 
Institute is supported in part by the Government of Canada through the Department of Innovation, Science and 
Economic Development and by the Province of Ontario through the Ministry of Colleges and Universities. This work 
was supported by a grant from the Simons Foundation (1034867, Dittrich).
\bibliographystyle{utphys}

\bibliography{refs}

\end{document}